\documentclass[a4paper,12 pt]{article}

\usepackage{amsmath} 
\usepackage{amsthm}
\usepackage{amssymb}
\usepackage{amsopn}

\begin{document}

\title{THE (UNFORTUNATE) COMPLEXITY OF THE ECONOMY}
\author{JP Bouchaud \footnote{Science \& Finance, Capital Fund Management,
6 Bd Haussmann, 75009 Paris, France}}

\maketitle

\begin{abstract}
This article is a follow-up of a short essay that appeared in Nature {\bf 455} 1181 (2008). It has become increasingly clear that the erratic dynamics of markets is mostly
endogenous and not due to the rational processing of exogenous news. I elaborate on the idea that {\it spin-glass} type of problems, where  the combination of competition and heterogeneities generically leads to long epochs of statis interrupted by crises and hyper-sensitivity to small changes of the environment, could be metaphors for the complexity of economic systems.  I argue that the most valuable contribution of physics to economics might end up being of methodological nature, and that simple models from physics and agent based numerical simulations, although highly stylized, are more realistic than the traditional models of economics that assume rational agents with infinite foresight and infinite computing abilities. 
\end{abstract}

The current direful crisis puts classical economics thinking under huge pressure. In theory, deregulated markets should be efficient, rational 
agents quickly correct any mispricing or forecasting error. Price faithfully reflect the underlying reality and ensure optimal allocation of resources. These ``equilibrated'' markets should be stable: crises can only be triggered by acute exogeneous disturbances, such as hurricanes, earthquakes or political upheavals, but certainly not precipitated by the market itself. This is in stark contrast with most financial crashes, including the latest one. The theory of economic equilibrium and rational expectations, as formalized since the 50's and 60's, has deeply influenced scores of decision-makers high up in government agencies and financial institutions. Some of them are now ``in a state of shocked disbelief'', as Alan Greenspan himself declared when he recently admitted that he had put too much faith in the self-correcting power of free markets and had failed to anticipate the self-destructive power of wanton mortgage lending. Economic theories turn out to have significant impact on our every-day life. The last twenty years of deregulation have been prompted by the argument that constraints of all kinds prevent the markets from reaching their supposedly perfect equilibrium, efficient state. The theory of Rational Expectations has now permeated into International Political Economics, Sociology, Law, etc.\footnote{There are rational expectation theories of marriage, drug addiction or obesity!} 

Unfortunately, nothing is more dangerous than dogmas donned with scientific feathers. The present crisis might offer an excellent occasion for a paradigm change, already called for in the past by yawing economists such as John Maynard Keynes, Alan Kirman or Steve Keen. They have forcefully highlighted the shortcomings and contradictions of classical economics, but progress has been slow.  Of course, it is all easier said than done, and the task looks so formidable that some economists argue that it is better to stick with the implausible but well corseted theory of perfectly rational agents rather than to venture into modelling the infinite number of ways agents can be irrational. So where should one start? What should be taught to students in order to foster, on the long run, a better grasp of the complexity of economic systems? Can physics really contribute to the much awaited paradigm shift? After twenty years or so of ``econophysics''\footnote{It seems adequate to define the first econophysics event as the Santa Fe conference in 1987, although the first scientific papers were written in the mid nineties, and the name econophysics coined by Gene Stanley in 1995.} and around 1000 papers in the arXiv, it is perhaps useful to give a personal bird's eye view of what has been achieved in that direction.  

Econophysics is in fact, at this moment in time, a misnomer since most of its scope concerns financial markets. To some economists, finance is of relatively minor importance and any contribution, even significant, can only have a limited impact on economics at large. I personally strongly disagree with this viewpoint: the recent events confirm that financial markets hiccups can cripple the entire economy. From a more conceptual point of view, financial markets represent an ideal laboratory for testing several fundamental concepts of economics, for example: Is the price really such that supply matches demand? Or: Are price moves primarily due to news?  The terabytes of data spitted out everyday by financial markets allows one (in fact compels one) to compare in detail theories with observations and the answers to both questions above seem to be clear no's.\footnote{On these points, see Joulin et al. (2008) for the minor role news seem to play in explaining large price jumps, and Bouchaud et al. (2008) for an extensive review on the inadequacy of the idea that supply and demand is cleared instantaneously in financial markets.}  

This proliferation of data should soon concern other spheres of economics and social science: credit cards and e-commerce will allow one to monitor consumption in real time and to test theories of consumer behaviour in great detail.
\footnote{For an interesting work in that direction, see Sornette et al (2004).}
So we must get prepared to deal with huge amounts of data, and to learn to scrutinize them with as little prejudice as possible, still asking relevant questions, starting from the most obvious ones -- those that need nearly no statistical test at all because the answers are clear to the naked eye -- and only then delving into more sophisticated ones. The very choice of the relevant questions is often sheer serendipity: more of an art than a science. That intuition, it seems to me, is well nurtured by an education in natural sciences, where the emphasis is put on mechanisms and analogies, rather than on axioms and theorem proving.  

Faced with a mess of facts to explain, Feynman advocated that one should choose one of them and try one's best to understand it in depth, with the hope that the emerging theory is powerful enough to explain many more observations. In the case of financial markets, physicists have been immediately intrigued by a number of phenomena described by power-laws. For example, the distribution of price changes, of company sizes, of individual wealth all have a power-law tail, to a large extent universal. The activity and volatility of markets have a power-law correlation in time, reflecting their intermittent nature, obvious to the naked eye: quiescent periods are intertwined with bursts of activity, on all time scales. Power-laws leave most economists unruffled (isn't it, after all, just another fitting function?), but immediately send physicists' imagination churning.  The reason is that many complex physical systems display very similar intermittent dynamics: velocity fluctuations in turbulent flows, avalanche dynamics in random magnets under a slowly varying external field, teetering progression of cracks in a slowly strained disordered material, etc. The interesting point about these examples is that while the exogeneous driving force is regular and steady, the resulting endogenous dynamics is complex and jittery.  In these cases, the non-trivial (physicists say {\it critical}) nature of the dynamics comes from collective effects: individual components have a relatively simple behaviour, but interactions lead to new, emergent phenomena. The whole is fundamentally different from any of its elementary sub-part.  Since this intermittent behaviour appears to be generic for physical systems with both heterogeneities and interaction, it is tempting to think that the dynamics of financial markets, and more generally of economic systems, does reflect the same underlying mechanisms. 

Several economically inspired models have been shown to exhibit these critical features. One is a transposition of the Random Field Ising Model (RFIM) to describe situations where there is a conflict between personal opinions, public information and social pressure.\footnote{See Sethna et al. (2001) for a general review on this model, and Galam \& Moscovici (1991) and Michard \& Bouchaud (2005) for some application to economics and social science.}  Imagine a collection of traders having all different a priori opinions, say optimistic (buy) or pessimistic (sell). Traders are influenced by some slowly varying global factors, for example interest rates, inflation, earnings, dividend forecasts, etc. One assumes no shocks whatsoever in the dynamics of these exogenous factors, but posits that each trader is also influenced by the opinion of the majority. He conforms to it if the strength of his a priori opinion is weaker than his herding tendency. If all agents made their mind in isolation (zero herding tendency) then the aggregate opinion would faithfully track the external influences and, by assumption, evolve smoothly. But surprisingly, if the herding tendency exceeds some finite threshold, the evolution of the aggregate opinion jumps discontinuously from optimistic to pessimistic as global factors only deteriorate slowly and smoothly. Furthermore, some hysteresis appears. Much as supersaturated vapor refusing to turn into liquid, optimism is self-consistently maintained. In order to trigger the crach, global factors have to degrade far beyond the point where pessimism should prevail. On the way back, these factors must improve much beyond the crash tipping point for global optimism to be reinstalled, again somewhat abruptly. Although the model is highly simplified, it is hard not to see some resemblance with all bubbles in financial history. The progressive reckoning of the amount of leverage used by banks to pile up bad debt should have led to a self-correcting, soft landing of global markets -- as the efficient market theory would predict. Instead, collective euphoria screens out all bad omens until it becomes totally unsustainable. Any small, anecdotal event or insignificant news is then enough to spark the meltdown.

The above framework also illustrates in a vivid way the breakdown of a cornerstone of classical economics, stigmatized in Alan Kirman's essay {\it Whom or what does the representative individual represent?} Much as in statistical physics or material science, one of the main theoretical challenges in economics is the micro/macro link.  How does one infer the aggregate behaviour (for example the aggregate demand) from the behaviour of individual elements? The representative agent theory amounts to replacing an ensemble of heterogeneous and interacting agents by a unique representative one -- but in the RFIM, this is just impossible: the behaviour of the crowd is fundamentally different from that of any single individual. 

Minority Games define another, much richer, family of models in which agents learn to compete for scarce resources.
\footnote{For a review and references, see Challet, Marsily and Zhang (2005).} 
A crucial aspect here is that the decisions of these agents impact the market: the price does not evolve exogenously but moves as a result of these decisions. A remarkable result is the existence, within this framework, of a genuine phase transition as the number of speculators increase, between a predictable market where agents can eke out some profit from their strategies, and an over-crowded market, where these profits vanish or become too risky. Around the critical point where predictability disappears and efficiency sets in, intermittent power-law phenomena emerge, akin to those observed on real stock markets. The cute point of this analysis is that there is a well-grounded mechanism to keep the market in the vicinity of the critical point:\footnote{A mechanism called Self-Organized Criticality by Per Bak (1996), which is assumed to take place in other situations potentially relevant to economics, for example evolution and extinction of species, which might have the evolution and extinction of companies.} less agents means more profit opportunities which attracts more agents, more agents means no profit opportunities so that frustrated agents leave the market.  

There are other examples in physics and computer science where competition and heterogeneities lead to interesting phenomena which could be metaphors of the complexity of economic systems: spin-glasses (within which spins interact randomly with one another), molecular glasses, protein folding, Boolean satisfiability problems, etc. In these problems, the energy (or the cost function) that must be minimized is an incredibly complicated function of the N degrees of freedoms (the spins, the position of the atoms of the protein, the Boolean variables). Generically, this function is found to display an exponential number (in N) of local minima. The absolute best one is (a) extremely hard to find: the best algorithms to find it take an exponential time in N; (b) only marginally better than the next best one; (c) extremely fragile to a change of the parameters of the problem: the best one can easily swap over to become the second best, or even cease abruptly to be a minimum. Physical systems with these ``rugged'' energy landscapes display very characteristic phenomena, extensively studied in the last twenty years, both experimentally and theoretically.\footnote{For a review see A. P. Young (1998)}  The dynamics is extremely slow as the system is lost amidst all these local minima; equilibrium is never reached in practice; there is intermittent sensitivity to small changes of the environment. There is no reason to believe that the dynamics of economic systems, also governed by competition and heterogeneities, should behave very differently -- at least beyond a certain level of complexity and interdependency.\footnote{The idea that spin-glass theory might be relevant to economics was originally suggested by Phil Anderson during the Santa Fe meeting {\it The economy as a complex evolving system} (1988).}

If true, this would entail a major change of paradigm:
\begin{itemize}
\item·First, even if an equilibrium state exists in theory it may be totally irrelevant in practice, because the equilibration time is far too long. As Keynes noted, {\it in the long run we are all dead}. The convergence to the Eden Garden of economic systems might not be hobbled by regulations but by their tug-induced complexity. One can in fact imagine situations where regulation could nudge free, competitive markets closer to an efficient state, which they would never reach otherwise.
\item·Second, complex economic systems should be inherently fragile to small perturbations, and generically evolve in an intermittent way, with a succession of rather stable epochs punctuated by rapid, unpredictable changes -- again, even when the exogenous drive is smooth and steady. No big news is needed to make markets lurch wildly, in agreement with recent empirical observations (see Joulin et al. 2008). Within this metaphor of markets, competition and complexity could be the essential cause of their endogenous instability.\footnote{In a recent beautiful paper, Marsili (2008) has shown how the increase of derivative products could drive the system close to an instability point, using concepts and methods quite similar to those of the Minority Game.}
\end{itemize}

The above models tell interesting stories but are clearly highly stylized and aim to be inspiring rather than convincing. Still, they seem quite a bit more realistic than the traditional models of economics that assume rational agents with infinite foresight and infinite computing abilities. Such simplifying caricatures are often made for the sake of analytical tractability, but many of the above results can in fact be established analytically, using statistical mechanics tools developed in the last thirty years to deal with disordered systems. One of the most remarkable breakthroughs is the correct formulation of a mean-field approximation to deal with interactions in heterogeneous systems. Whereas the simple Curie-Weiss mean-field approximation for homogenous systems is well known and accounts for interesting collective effects\footnote{The Curie-Weiss mean field theory was first used in an economic context by Brock and Durlauf (2001).}, its heterogeneous counterpart is far subtler and has only been worked out in detail in the last few years.\footnote{For a thorough review, see M\'ezard  \& Montanari (2009).} It is a safe bet to predict that this powerful analytical tool will find many natural application in economics and social sciences in the years to come. 

As models become more realistic and hone in on details, analytics often has to give way to numerical simulations. The situation is now well accepted in physics, where numerical experimentation has gained a respectable status, bestowing us with a {\it telescope of the mind,} (as beautifully coined by Mark Buchanan) {\it multiplying human powers of analysis and insight just as a telescope does our powers of vision}. Sadly, many economists are still reluctant to recognize that numerical investigation of a model, although very far from theorem proving, is a valid way to do science. Yet, it is a useful compass to venture into the wilderness of irrational agent models: try this behavioural rule and see what comes out, explore another assumption, iterate, explore. It is actually surprising how easily these numerical experiments allow one to qualify an agent-based model as potentially realistic (and then one should dwell further) or completely off the mark.\footnote{For reviews, see Goldberg \& Janssen (2005) and Lux (2008).} What makes this expeditious diagnosis possible is the fact that for large systems details do not matter much -- only a few microscopic features end up surviving at the macro scale. This is a well-known story in physics: the structure of the Navier-Stokes equation for macroscopic fluid flow, for example, is independent of all molecular details. The present research agenda is therefore to identify the features that explain financial markets and economic systems as we know them. This is of course still very much of an open problem, and simulations will play a central role. The main bet of econophysics is that competition and heterogeneity, as described above, should be the marrow ingredients of the final theory. 

A slew of other empirical results, useful analytical methods and numerical tricks have been established in the 15 active years of econophysics, which I have no space to review here.\footnote{ Let me quote in particular models of wealth distribution, market microstructure and impact of trades, exactly solvable stochastic volatility models, path integrals, multifractal random walks or random matrix theory. For more exhaustive reviews, see (among others): Bouchaud \& Potters (2004), Yakovenko (2007), Lux (2008).}  But in my opinion the most valuable contribution of physics to economics will end up being of methodological nature. Physics has its own way to construct models of reality based on a subtle mixture of intuition, analogies and mathematical spin, where the ill-defined concept of plausibility can be more relevant than the accuracy of the prediction. Kepler's ellipses and Newton's gravitation were more plausible than Ptolemy's epicycles, even when the latter theory, after centuries of fixes and stitches, was initially more accurate to describe observations. When Phil Anderson first heard about the theory of Rational Expectations in the famous 1987 Santa Fe meeting, his befuddled reaction was: {\it You guys really believe that?} He would probably have fallen from his chair had he heard Milton Friedman's complacent viewpoint on theoretical economics: {\it In general, the more significant the theory, the more unrealistic the assumptions}. Physicists definitely want to know what an equation means in intuitive terms, and believe that assumptions ought to be both plausible and compatible with observations. This is probably the most urgently needed paradigm shift in economics.

\section*{References}

P. W. Anderson, K. Arrow, D. Pines, {\it The Economy as an Evolving Complex System}, Addison-Wesley (1988) 

P. Bak, {\it How Nature Works: The Science of Self-Organized Criticality}, New York: Copernicus (1996)

J.-P. Bouchaud and M. Potters, {\it Theory of Financial Risks and Derivative Pricing},  Cambridge University Press, 2003.

J.-P. Bouchaud, J. D. Farmer, F. Lillo,  {\it How Markets Slowly Digest  Changes in Supply and Demand}, in: Handbook of 
Financial Markets: Dynamics and Evolution, North-Holland, Elsevier, 2009.

W. Brock, S. Durlauf, {\it Discrete Choice with social interactions}, Review of Economic Studies, 68, 235 (2001).

M. Buchanan, {\it This Economy does not compute}, New York Times, Oct. 1st 2008.

D. Challet, M. Marsili, Y.C. Zhang, {\it Minority Games}, Oxford University Press (2005)

J. D. Farmer, J. Geanakoplos, {\it The virtues and vices of equilibrium and the future of financial economics}, e-print arXiv:0803.2996 (2008)

S. Galam, S. Moscovici, {\it Towards a theory of collective phenomena: Consensus and attitude changes in groups},  Euro.  J.  Social  Psy. 21, 49 (1991)

R. Goldstone, M. Janssen, {\it Computational Models of collective behaviour}, Trends in Cognitive Science, 9, 424 (2005)

A. Joulin, A. Lefevre, D. Grunberg, and J.-P. Bouchaud, {\it Stock price jumps: News and volume play a minor role}, arXiv:0803.1769; Wilmott Magazine, November 2008.

S. Keen, {\it Debunking Economic}, Pluto Press (2000)

A. Kirman, {\it What or whom does the representative individual represent?}, Journal of Economic Perspectives, 6, 117 (1992)

T. Lux, {\it Applications of Statistical Physics in Finance and Economics}, to appear in Handbook of Research on Complexity (2008)

R. N. Mantegna, H. E. Stanley, {\it An Introduction to Econophysics: Correlations and Complexity in Finance}, Cambridge University Press (1999)

M. M\'ezard, A. Montanari, {\it Information, Physics \& Computation}, Oxford University Press (2009)

Q. Michard, J.-P. Bouchaud,  {\it Theory of collective opinion shifts:  from smooth trends to abrupt swings}, Eur. J. Phys. B 47, 151 (2005)

M. Marsili, {\it Eroding Market Stability by Proliferation of Financial Instruments} (2008):  http://ssrn.com/abstract=1305174

J. Sethna, K. Dahmen, C. Myers, {\it Crackling Noise}, Nature, 410, 242 (2001)

D. Sornette, F. Desch\^atres, T. Gilbert, Y. Ageon, {\it Endogeneous vs exogeneous shocks in complex systems: an empirical test using book sales ranking}, Phys. Rev. Lett. 93, 228701 (2004).

V. Yakovenko, {\it Statistical Mechanics approach to Econophysics}, arXiv:0709.3662 (2007), to appear in Encyclopedia of Complexity and System Science

A. P. Young, {\it Spin glasses and Random Fields}, World Scientific (1997)

\end{document}